\documentclass[twocolumn,12pt,tighten,twocolappendix]{aastex631}
\usepackage{amsmath,amssymb}

\usepackage{lipsum}
\usepackage{graphicx}
\usepackage{subfigure}
\renewcommand{\l}{\left}
\renewcommand{\r}{\right}

\hypersetup{linkcolor=blue,citecolor=blue,filecolor=cyan,urlcolor=magenta}
\usepackage{hyperref}
\usepackage{cleveref}
\usepackage{comment}

\usepackage[normalem]{ulem}

\usepackage[normalem]{ulem}
\begin{document}

\title{Cost of inferred nuclear parameters towards the f-mode dynamical tide in binary neutron stars}


\author[0000-0002-2526-1421]{Bikram Keshari Pradhan}
\email{bikramp@iucaa.in}
\author[0000-0001-9848-9905]{Tathagata Ghosh}
\author[0000-0001-8129-0473]{Dhruv Pathak}
\author[0000-0002-0995-2329]{Debarati Chatterjee}
\affiliation{Inter University Centre for Astronomy and Astrophysics, Pune, Maharastra, 411007, India}


\begin{abstract}
 Gravitational Wave (GW) observations from Neutron Stars (NS) in a binary system provide an excellent scenario to constrain the nuclear parameters. The investigation of ~\cite{Pratten2022} has shown that the ignorance of f-mode dynamical tidal correction in the GW waveform model of the binary neutron star (BNS) system can lead to substantial bias in the measurement of NS properties and NS equations of state (EOS). In this work, we investigate the bias in the nuclear parameters resulting from the ignorance of dynamical tidal correction. In addition, this work demonstrates the sensitivity of the nuclear parameters and the estimated constraints on nuclear parameters and NS properties from future GW observations.  We infer the nuclear parameters from GW observations by describing the NS matter within the relativistic mean field model. For a population of GW events, we notice that the ignorance of dynamical tide predicts a lower median for nucleon effective mass ($m^*$) by $\sim6\%$ compared to the scenario when dynamical tidal correction is considered. Whereas at a 90\% credible interval(CI), $m^*$ gets constrained up to $\sim 5\%$ and $\sim 3\%$ in A+ (the LIGO-Virgo detectors with a sensitivity of 5th observing run) and Cosmic Explorer (CE) respectively. We also discuss the resulting constraints on all other nuclear parameters, including compressibility, symmetry energy, and slope of symmetry energy,   considering an ensemble of GW events. We do not notice any significant impact in analyzing nuclear parameters other than $m^*$ due to the ignorance of f-mode dynamical tides.

\end{abstract}

\section{Introduction} \label{sec:intro}
Neutron stars (NS) provide a natural laboratory to explore the ultra-high dense matter under extreme conditions, such as high magnetic field and rotation, that are beyond the reach of the terrestrial experiments~\citep{Lattimer2021}. The NS macroscopic properties relate to the microscopic description of the NS matter via the pressure density relationship, more commonly known as the equation of state (EOS).   The measurement of NS observables such as mass ($M$) and radius ($R$)  with electromagnetic observations~\citep{Riley_2021,Miller_2021}  at multiple wavelengths have significantly improved our understanding regarding the NS interior and the NS EOS. Additionally, the binary neutron star (BNS) mergers are sources of gravitational waves (GW). In a BNS merger, the tidal deformation of the NSs depends upon the NS interior, and hence, the measurement of the tidal deformability parameter is used to constrain the NS EOS in combination with mass measurements from the same binary. The detection of  BNS events GW170817 ~\citep{AbbottPRL119,AbbottPRL121,AbbottAJL848,AbbottPRX2019,Abbott_GWTC1} and GW190425 ~\citep{Abbott_2020,AbbottGWTC2} by the LIGO-Virgo collaboration ~\citep{LIGOScientific:2014pky,VIRGO:2014yos} have been widely used to understand the NS interior~\citep{Most2018,Biswas2021,Bauswein_2017,Essick2021,Huth2022,Annala2018,Fattoyev2018} and have opened a new window in the multimessenger astronomy.

The tidal deformation of the stars in a BNS contributes to the phase of the GW signal. The dominant contribution arises from the electric type quadrupolar adiabatic tidal deformation appearing first at $5$ post-Newtonian (PN) order~\citep{Hinderer2010}, and this has been used to infer the NS properties from the detected BNS candidates. A significant amount of effort has been made to develop an accurate waveform model, including different corrections such as the inclusion of higher order electric and magnetic multipolar tidal corrections~\citep{Hinderer2010,Bini2012,Damour2010,Landry2015,Banihashemi2020,Poisson2020,Henry2020,Mandal:2023hqa} and the dynamical tidal correction due to resonant/non-resonant excitation of NS oscillation modes~\citep{Hinderer2016,Steinhoff2016,Schmidt2019,Kuan2022,Pnigouras2022,Gamba2022,Gupta2021,Gupta2023,Mandal2023,Abac:2023ujg}. Although, for the detected events GW170817 and/or GW190425, there is a minor impact on the inferred NS properties due to the additional corrections of multipolar tidal effects~\citep{Godzieba2021,Pradhan2023,Narikawa2023} or due to the consideration of the dynamical corrections~\citep{Pratten2020,Gamba2022,Pradhan2023,Abac:2023ujg}, their importance will be significant for future events accessible with increasing sensitivity of the current GW detectors or even with the next generation detectors. 
Furthermore, GW models, including dynamical tides, show better agreement with the numerical relativity simulations~\citep{Hinderer2016,Steinhoff2016,Andersson2019,Schmidt2019}. The recent investigation of~\cite{Pratten2022} suggests that the ignorance of f-mode dynamical tides can sufficiently bias the measurement of NS tidal deformability, leading to the biased inference of the EOS.

The simultaneous measurements of mass and tidal deformability of the NSs from GW measurements play a crucial role in improving the uncertainty in the nuclear physics~\citep{Margueron2020,Ghosh2022,Ghosh:2022lam,Biswas2021,Lattimer2021,PRADHAN2023122578,Mondal2023,Huang2023,Mikhail2023,Traversi2022,Pradhan:2023zor,Patra:2022lds,Imam:2021dbe}. As the ignorance of the f-mode dynamical tides substantially affects the measurement of NS EOS and NS properties, this leads us to question how much bias one would expect on the nuclear parameters constrained using GW observations due to ignorance of f-mode dynamical tides. In the context of NS physics, the EOS models based on the inclusion of interaction among the  NS constituents are classified into two main subgroups: (i) microscopic or {\it ab-initio}~\citep{OertelRMP,Sabatucci2022} and (ii) phenomenological models described by effective theories with parameters fitted to reproduce saturation nuclear parameters and/or nuclear properties~\citep{MACHLEIDT19871}. As our main concern is to study the bias in the nuclear parameters, we adopt the phenomenological relativistic mean field (RMF) model whose model parameters are calibrated to nuclear parameters~\citep{Hornick,chen2014}.

The recent study by \cite{Iacovelli:2023nbv} focuses on constraining nuclear parameters based on forthcoming GW observations. However, contrary to the meta-model approach used in  ~\cite{Iacovelli:2023nbv}, our work differs by using the RMF description of the NS EOS. The comprehensive investigation in the study by \cite{Iacovelli:2023nbv} employs the Fisher matrix analysis for GW parameter inference. However, a thorough comparison of the Fisher matrix analysis with the complete Bayesian analysis in ~\cite{Iacovelli:2023nbv} reveals that the Fisher matrix approach falls short in accurately estimating the errors associated with NS properties in certain cases. As our analysis involves posterior distributions of individual tidal parameters, we perform the Bayesian parameter estimation instead of the Fisher matrix analysis. Furthermore, the recent work from ~\citep{Walker2024} discusses the recession constraints on the NS EOS with the next-generation gravitational wave detector using the spectral decomposition method~\citep{spectral} to describe the NS EOS.

In this work, we try to translate the bias in the measured NS properties from GW events due to ignorance of f-mode dynamical tide in the nuclear parameters in a Bayesian formalism. The work is organized in the following way. In Sec.~\ref{sec:method}, we discuss the methodology of our work, including the choice of GW events, detector configurations, and Bayesian formalism. We discuss our results in Sec.~\ref{sec:results} and conclude our findings in Sec.~\ref{sec:conclusion}.

\section{Methodology}\label{sec:method}
To investigate the impact of f-mode dynamical tides, we model the GW  with the frequency domain \texttt{TaylorF2} waveform model including  3.5PN point particle phase, adiabatic tidal effects up to 7.5PN order accounting for the impact from magnetic deformation ($\Sigma_2$), and the octupolar tidal deformability ($\Lambda_3$)~\cite{Henry2020}: we represent this waveform as ~$\rm TF2_{AT}$. Additionally, we include the ready-to-use quadrupolar f-mode dynamical tidal correction from ~\cite{Schmidt2019} to $\rm TF2_{AT}$ waveform model and represent this as $\rm TF2_{DT}$. 

{As discussed in~\cite{Pratten2022}, the impact of the dynamical tide is  significant for the $5$th observation run (O5) of LIGO-Virgo detectors~\citep{2015CQGra..32g4001L, 2015CQGra..32b4001A} and even for next-generation detectors. So, we focus on the GW events detected by LIGO-Virgo detectors with A+ sensitivity (as anticipated for the $5$th observation run)~\citep{Aplus,Aplus_sensitivity} and the next-generation detector, cosmic explorer (CE)~\citep{CE,CE2,CE3,CE_Sensitivity}.} We consider uniform mass distribution between $M_{\rm min} = 1\ M_{\odot}$ and maximum mass $M_{\rm max}$ for each component mass. 
The choice of $M_{\rm min}$ is consistent with the predicted lower bound of NS mass from plausible supernova formation channels~\citep{2012ApJ...749...91F, Woosley:2020mze}.
The maximum mass is determined from the injected EOS.
The other relevant properties, such as the tidal deformability parameters of electric type $\Lambda_{\ell=\{2,3\}}$, magnetic type $\Sigma_2$, and f-mode angular frequency $\omega_2$ are assigned to the NSs assuming a particular nuclear parametrization~\footnote{The particular parameterization is considered for the RMF model given in section~\ref{sec:EOS} (for which we are aiming to constrain the nuclear parameters), and it also satisfies the state-of-the-art constraints such as Chiral Effective Field Theory and current astrophysical data. } given in the injection column of ~\cref{tab:prior_posterior} with the EOS model described in ~\cite{Hornick}.
The sources are distributed isotropically over the sky and within the luminosity distance ($d_L$) of $200$ Mpc following Madau-Dickinson star formation rate~\citep{2014ARA&A..52..415M}. We assume Planck15 cosmology~\citep{2016A&A...594A..13P} in this entire work.
For the injection and recovery studies discussed in ~Sec.~\ref{sec:injection}, we ignore the spin of individual NSs.
Furthermore, considering a local merger rate of $\sim 450$ Gpc$^{-3}$ yr$^{-1}$~\citep{2023PhRvX..13a1048A}, we expect $\sim 13$ BNS events per year. 

Next, we perform parameter estimation using the dynamic nested sampler \texttt{dynesty} \cite{Dynesty} as implemented in the parameter estimation package ~\verb+bilby_pipe+~\citep{bilby_2019} for individual BNS events. We consider uniform prior over detector-frame component-masses ($m_{1,2}^{z}=m_{1,2}(1+z)$, where $m_{1,2}$ are source-frame masses at redshift $z$) and the corresponding tidal deformabilities ($\Lambda_{1,2}$). We also assume a power law prior for luminosity distance ($d_{L}^{2}$) and isotropic prior for inclination angle. For the additional parameters, such as $\Lambda_3,\Sigma_2,\  \text{and}\  \omega_2$, we use the universal relations from ~\cite{Pradhan2023}. 
{The remaining BNS parameters (such as sky position) are kept fixed at their injected values to reduce the computational cost. However, one should ideally use broad enough priors for all the source parameters. This is a simplification in this work that should be improved in the future to investigate the uncertainties and systematics in inferred EoS parameters.} 
The posteriors thus obtained will be utilized further to constrain the EOS parameters $\bm{\mathcal{E}}$ as follows:

\begin{equation} \label{bayesian_EOS}
    p(\bm{\mathcal{E}} \mid \{d\}) \propto \pi(\bm{\mathcal{E}}) \prod_{i} \int \mathcal{L}(d_{i} \mid \bm{\theta}_{i}) \pi_{\rm pop} (\theta_{i} \mid \bm{\mathcal{E}}) d\bm{\theta}_{i} 
\end{equation}
where, $\bm{\theta}=\{m_{1}, m_{2}, d_{L}, \Lambda_{1}, \Lambda_{2}\}$ represent the source parameters of the individual source, denoted by the subscript $i$. Since we are only interested in inferring EOS parameters from $m-\Lambda$ relation and fixing the population, the selection effect has not been included in Eq.~\eqref{bayesian_EOS}. In Eq.~\eqref{bayesian_EOS}, the semi-marginalized likelihood $\mathcal{L}$ has been calculated by dividing the priors $\pi_{\rm PE}(\bm{\theta})$ from the corresponding posteriors $p(\bm{\theta} \mid d)$ while performing the  parameter estimation for NS properties using ~\verb+bilby_pipe+:

\begin{equation}
    \mathcal{L}(d \mid \bm{\theta}) \propto \frac{p(\bm{\theta} \mid d)}{\pi_{\rm PE}(\bm{\theta})}
\end{equation}

We have implemented the Gaussian kernel density estimator from ~\verb+Statsmodel+~\citep{seabold2010statsmodels} to estimate the likelihood $\mathcal{L}$ from the posterior samples of source parameters.
Finally, the EOS parameters have been estimated by using the nested sampler~\verb+Pymultinest+~\citep{2014A&A...564A.125B}, as shown in Eq.~\eqref{bayesian_EOS}.



\subsection{Equations of State: RMF Model}\label{sec:EOS}

 The pressure density relationship: $p=p(\epsilon)$, referred to as the EOS, plays a vital role in relating the microscopic behavior of NS matter and the properties of the NS. We describe the NS matter with the relativistic mean field (RMF) model, which is a  phenomenological model where the model parameters are calibrated to the nuclear parameters saturation data.  In RMF theory, the Lagrangian density describes the interaction between baryons through the exchange of mesons: the scalar-isoscalar ($\sigma$), vector-isoscalar ($\omega$), vector-isovector ($\rho$) mesons as given in Eq.~\eqref{eqn:lagr}. 
\begin{eqnarray}
     \mathcal{L} &=&\sum_N  \bar{\psi}_{N}  (i\gamma^{\mu}\partial_{\mu}-m +g_{\sigma}\sigma-g_{\omega B}\gamma_{\mu}\omega^{\mu}-\frac{g_{\rho}}{2}\gamma_{\mu} \vec{\tau}\vec{\rho}^{\mu})\psi_{N} \nonumber \\
     &+&\frac{1}{2}  (\partial_{\mu} \sigma \partial^{\mu}\sigma - m_{\sigma}^2 {\sigma}^2) 
     -\frac{1}{3}b m   (g_{\sigma} \sigma)^3+\frac{1}{4}c   (g_{\sigma} \sigma)^4 \nonumber\\
     &+&\frac{1}{2}m_{\omega}^2 \omega_{\mu}\omega^{\mu}-\frac{1}{4} \omega_{\mu \nu}\omega^{\mu \nu} \nonumber \\
     &-&\frac{1}{4}  (\vec{\rho}_{ \mu \nu}.\vec{\rho}^{\mu \nu}-2 m_{\rho}^2 \vec{\rho}_{\mu}\vec{\rho}^{\mu})+\Lambda_{\omega} (g_{\rho}^2  \vec{\rho}_{\mu} \vec{\rho}^{\mu}) \  (g_{\omega}^2 \omega_{\mu}\omega^{\mu})  
     \label{eqn:lagr}
\end{eqnarray}
where $\Psi_{N}$ is the Dirac field of the nucleons N, $m$ is the vacuum nucleon mass, while $\gamma^{\mu}$ and $\vec{\tau}$ are the Dirac and Pauli matrices respectively. $\sigma$, $\omega$, $\rho$ denote the meson fields, with isoscalar coupling constants  $g_{\sigma}$, $g_{\omega}$, isovector coupling $g_{\rho}$ and mixed $\omega - \rho$ coupling $\Lambda_{\omega}$. $b$ and $c$ represent the scalar meson self-interaction.

 The isoscalar set of couplings $g_{\sigma}$, $g_{\omega}$, $b$ and $c$ are determined by fixing the saturation density $n_0$, the binding energy per nucleon $E_{sat}$, the incompressibility coefficient $K$ and the effective nucleon mass $m^*= m_N - g_{\sigma}\sigma$ at saturation, $m_N$ being the mass of the nucleon at the saturation. On the other hand, isovector couplings $g_{\rho}$ and $\Lambda_{\omega}$ are determined as a function of the symmetry energy $J$ and the slope of the symmetry energy at saturation $L$. Complete descriptions regarding obtaining the energy density and pressure in the mean-field limit can be found at \cite{Hornick}. Hence, given a set of nuclear parameters, $\bm{\mathcal{E}}=\{n_0, E_{sat},K,m^*,L,J\}$, a unique EOS can be obtained following the methodology given in ~\cite{Hornick}. We fix the low density EOS to that of ~SLy~\citep{Gulminelli2015} and stitch it to the core EOS at a density $<0.5n_0$, making the EOS thermodynamically stable and also satisfying the causality condition (i.e., speed of sound $\leq1$).
 
 \section{Results}\label{sec:results}
 
\subsection{GW170817}
Several investigations have been conducted to determine the impact of the dynamical tidal correction for the event GW170817~\citep{Pratten2020,Gamba2022,Pradhan2023}. 
Even though there is no statistically significant support (based on the Bayes factor comparison) for the inclusion of dynamical tidal corrections over adiabatic tides for the GW170817, the bias on the recovered tidal parameters cannot be disregarded~\citep{Gamba2022,Pradhan2023}. 
This can be explained by looking at the relatively low sensitivity of the GW detector in the region where the dynamical tidal correction appears significant ($\geq 800 $ Hz, \citep{Williams2022}).

We perform the Bayesian parameter estimation of GW170817 strain data~\citep{RICHABBOTT2021}\footnote{\url{https://www.gw-openscience.org/events/GW170817/}} with the noise curve given in~\citep{Abbott_GWTC1}, using the dynamic nested sampler \texttt{dynesty}~\citep{Dynesty} as implemented in the parameter estimation package \texttt{bilby\_pipe} ~\citep{bilby_2019}. In addition to the inspiral-only frequency domain TaylorF2 ($\rm TF2_{AT}$ and $\rm TF2_{DT}$) waveform models described in Sec.~\ref{sec:method}, we also consider the spin orbit~\citep{Boh_2013} and spin-spin interaction corrections~\citep{Mishra2016} to the waveform model. Further details can be found in ~\cite{Pradhan2023}~\footnote{We have not used the ```relative binning" methodology here to find the posterior of the BNS event GW170817, as done in our previous work ~\cite{Pradhan2023}.}.
\begin{figure}
    \centering
    \includegraphics[width=\linewidth]{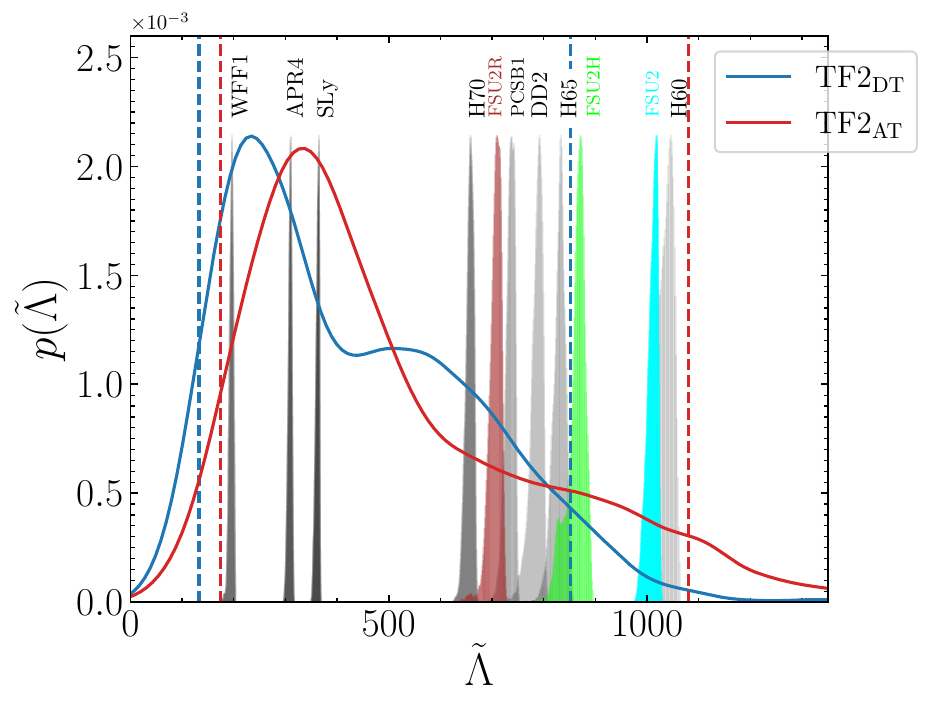}
    \caption{The probability distribution of $\tilde{\Lambda}$ resulting from the BNS event GW170817 with (blue) and without (red) consideration of f-mode dynamical tidal correction. The 90\% symmetric credible interval has also been shown with dashed lines. The distribution of $\tilde{\Lambda}$ for a few EOS models has also been displayed. The probability distributions of $\tilde{\Lambda}$ for representative EOS models are scaled to display them in one figure for better comparison.}
    \label{fig:gw170817_ltilde}
\end{figure}

 We show the posterior probability distribution of the reduced tidal deformability parameter $\tilde{\Lambda}$ (see relation (5) from~\cite{Wade2014} for the definition) resulting from waveform models $\rm TF2_{AT}$ and $\rm TF2_{DT}$  in \cref{fig:gw170817_ltilde}. Additionally, \cref{fig:gw170817_ltilde} displays the distribution of $\Tilde{\Lambda}$ for a few representative EOSs: WFF1~\citep{WFF1}, APR4~\citep{APR}, SLy4~\citep{CHABANAT1998231,Gulminelli2015,DANIELEWICZ200936}, PCSB1~\citep{PRADHAN2023122578}, DD2~\citep{DD2} and the FSU2, FSU2H, FSU2R models from ~\citep{FSU,Tolosfsu,Negreiros_2018,Grill2014}. 
 The aforementioned realistic EoSs are taken either from CompOSE database  ~\citep{Compose1,Compose2,Compose3} or from LALSimulation ~\citep{lalsuite}.
  We have also shown the  distributions of $\Tilde{\Lambda}$ corresponding to the Hornick parametrizations~\citep{Hornick} H60, H65, and H70, with different values for  $m^*/m_N=0.6, 0.65$, and 0.70, respectively~\footnote{The other nuclear parameters are fixed at $n_0=0.15 \ \rm fm^{-3}$, $E_{sat}=-16$MeV, $K=240$ MeV, $L=60$ MeV, $J=32$ MeV}. Though the distributions of $\Tilde{\Lambda}$ for RMF parametrized EOSs FSU2H, FSU2, H60 successfully fall within the $90\%$ symmetric credible interval (CI) of $\Tilde{\Lambda}$ posterior recovered using $\rm TF_{AT}$, it does not fit within the $90\%$ symmetric credible interval (CI) of $\Tilde{\Lambda}$ recovered using $\rm TF_{DT}$. This suggests that the nuclear parameters constrained using the GW detection can be affected by considering f-mode dynamical tidal corrections.

 We perform the Bayesian analysis to constrain the EOS parameters described in Sec.~\ref{sec:method}. We consider uniform prior for each nuclear parameter $x$ ranging from $x_l$ to $x_u$, similar to the values given as the truncation limits in the prior column of ~\cref{tab:prior_posterior}. 
 For GW170817, considering a uniform prior in nuclear parameters, we find that only $m^*$ and $L$ get better constrained, but no noticeable constraint on the other nuclear parameters is observed. 
 We compile the recovered distribution of $m^*$ and $L$ in ~\cref{fig:gw170817_mstar_lsym} of Appendix~\ref{app:mstar_LwithGW170817}. For GW170817, the consideration of $\rm TF2_{DT}$ GW model does not impact the posterior of nuclear parameters that were recovered with  $\rm TF2_{AT}$ GW model. 
 However, we notice a smaller value for the lower bound of $m^*$ resulting in $\rm TF2_{DT}$ GW model compared to what was recovered with $\rm TF2_{AT}$. This can be explained by looking at the higher upper limit of $\tilde{\Lambda}$ with $ \rm TF2_{AT}$, demanding stiffer EOSs and resulting in a smaller lower bound on $m^*$.

\subsection{Injection Studies}\label{sec:injection}
 The impact of f-mode dynamical tide becomes significant with the increasing sensitivity of the current detectors or even for the next generation detectors~\cite{Pratten2022}. We consider two different detector network configurations. The first one consists of two LIGO detectors ( LIGO-Hanford(H) and LIGO-Livingston (L)) with the A+ design sensitivity, as anticipated for the fifth observing run (O5) and the Virgo detector(V) operating at the
low-limit sensitivity~\citep{Aplus_sensitivity} (similar to that of ~\cite{Pratten2022}). The other network consists of two Cosmic Explorer (CE) detectors~\citep{CE,CE2}, located at Hanford and Livingston, operating at their designed sensitivity~\citep{CE_Sensitivity}. Since the simulated sources are at low redshifts, all the events have a significant network signal-to-noise ratio (SNR) of $\geq$ 30 and $\geq$ 200 for A+ and CE configuration, respectively.

 To study the bias due to ignorance of f-mode dynamical tides, we obtain the posteriors of GW parameters using both  $\rm TF2_{AT}$ and $\rm TF2_{DT}$  waveform models individually while keeping the injection waveform to $\rm TF2_{DT}$, i.e., the dynamical correction is always considered during the injection. We perform the parameter estimation for the GW parameters and the nuclear parameters for the simulated events described in Sec.~\ref{sec:method}. We employ a truncated Gaussian prior for each of the nuclear parameters ($x$) with a mean ($\mu_x$) fixed to the parameters of injected PCSB0 EOS model and the standard deviation ($\sigma_x$) inspired from \citet{Margueron2018}. We truncate  each normal distribution for  parameter $x$ within the range, defined by the minimum value $x_l$ and  the maximum value $x_u$, i.e, the prior of $x$ is taken as ~$\pi(x)=\mathcal{N}(\mu_x,\sigma_x) \ T[x_l,x_u]$.

\begin{figure*}
    \centering
    \includegraphics[width=\linewidth]{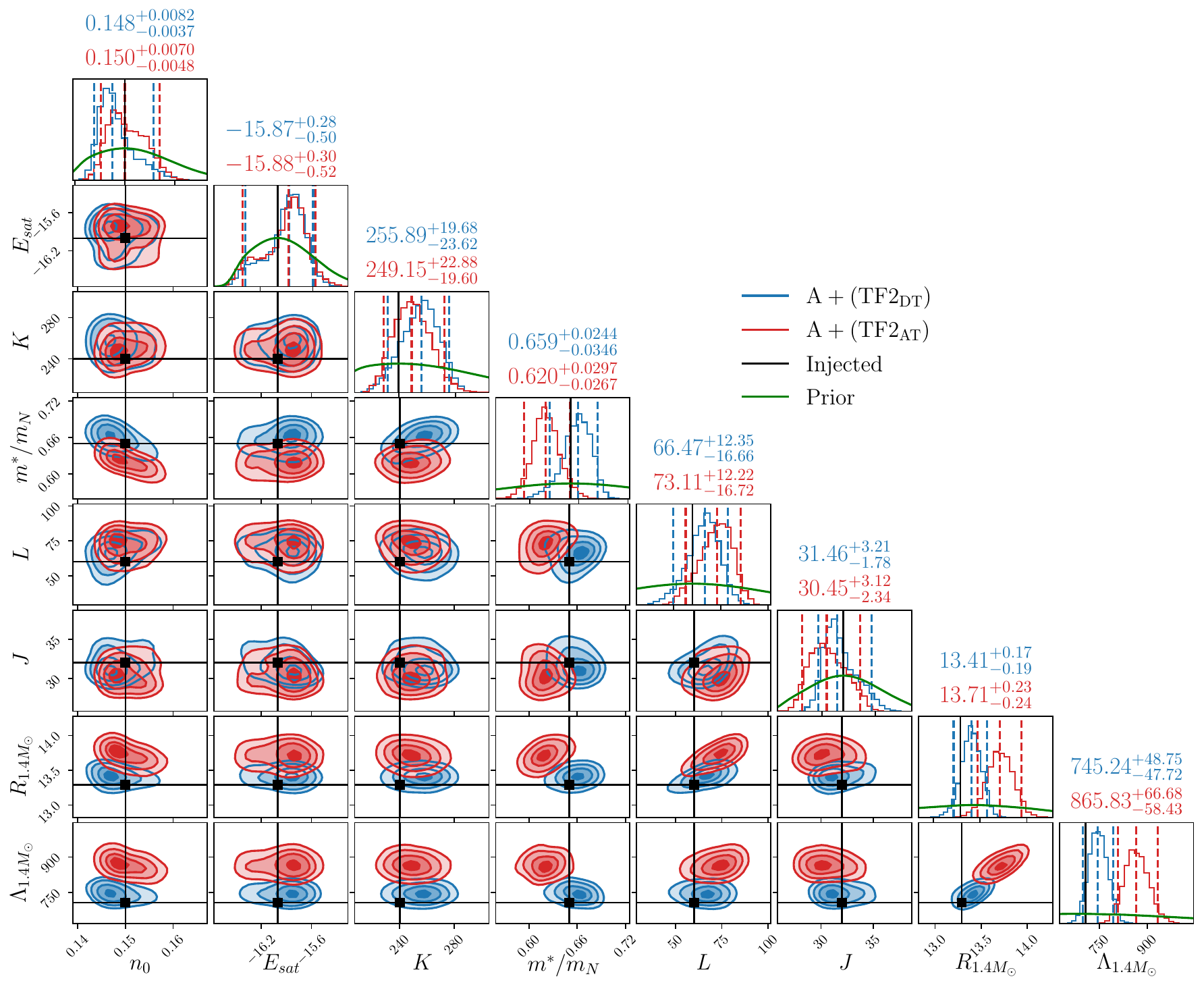}
    \caption{Joint and marginalized posterior distribution of nuclear parameters resulting from 13 BNS events in A+ configuration with (blue) and without (red) consideration of f-mode dynamical correction in the  GW model during recovery of NS properties from simulated GW events, while injections are done with the inclusion of dynamical tidal correction. The median, the upper, and the lower bound correspond to the symmetric 90\% credible interval are also mentioned. The injection values are shown with black lines, and the priors are displayed in green. The units of nuclear parameters are mentioned in ~\cref{tab:prior_posterior} and $R_{1.4M_{\odot}}$ is in units of km.}
    \label{fig:Nuclear_aplus}
\end{figure*}

\begin{table*}
    \centering\setlength\tabcolsep{.44em}
    \begin{tabular}{|c|c|c|c|c|}
    \hline
    \hline
        Parameter& Injection & Prior & \multicolumn{2}{c|}{Posterior}   \\
         \cline{4-5}
         & & &    A+  &  CE-CE \\
          &   & & $\rm TF2_{DT}$($\rm TF2_{AT}$) & $\rm TF2_{DT}$($\rm TF2_{AT}$ ) \\
          \hline
          \hline
         $n_0 \rm{[fm^{-3}]}$&0.15  &$\mathcal{N} (0.15,0.006) T[0.14,0.18] $ &   $0.148^{+0.0082}_{-0.0037}$ &$0.148^{+0.0023}_{-0.0020}$\\
         & & &  $\l(0.150^{+0.0070}_{-0.0048}\r)$ &$\l(0.147^{+0.0022}_{-0.0017}\r)$ \\
         \hline
         $E_{sat}\rm{[MeV]}$&-16.0   &$\mathcal{N} (-16,0.40)\  T[-16.5,-15] $ &  $-15.87^{+0.28}_{-0.50}$  & $-15.69^{+0.24}_{-0.46}$   \\
         & & &  $\l(-15.88^{+0.30}_{-0.52}\r)$ &$\l(-15.71^{+0.18}_{-0.34}\r)$  \\
         \hline 
         $K\rm{[MeV]}$&240 &$\mathcal{N} (240,50)\  T[200,355] $ &   $256^{+20}_{-24}$ & $250^{+20}_{-29}$\\
         & & &  $\l(249^{+23}_{-20}\r)$ &$\l(253^{+15}_{-25}\r)$  \\
         \hline
         $m^*/m_N$ &0.65&$\mathcal{N} (0.65,0.09)\  T[0.4,0.9] $ &   $0.659^{+0.024}_{-0.035}$ & $0.657^{+0.014}_{-0.015}$ \\
         & & &   $\l(0.620^{+0.03}_{-0.027}\r)$ & $\l(0.620^{+0.011}_{-0.014}\r)$ \\
         \hline 
         $L\rm{[MeV]}$&60 &$\mathcal{N} (60,35)\  T[1,140] $ &   $66^{+12}_{-17}$ & $58^{+7}_{-6}$ \\
         & & &   $\l(73^{+12}_{-17}\r)$ &$\l(62^{+9}_{-6}\r)$\\
         \hline
         $J\rm{[MeV]}$&32 &$\mathcal{N} (32,3)\  T[26,39] $ &  $31.46^{+3.21}_{-1.78}$ & $32^{+1.65}_{-1.58}$ \\
          & & &  $\l(30.45^{+3.12}_{-2.34}\r)$ & $\l(31.2^{+1.81}_{-2.46}\r)$ \\
         \hline

        \hline \hline
    \end{tabular}
    \caption{The median and 90\% symmetric credible interval of the recovered posterior of nuclear parameters resulting from different scenarios considered in this work. The parameters of injected EOS are also tabulated in the injection column. The priors considered in the work are truncated Gaussian distributions with mean at the injected value and deviations are inspired from ~\citet{Margueron2018}. We also truncate the distribution of nuclear parameters in a range that includes the minimum and maximum values, resulting from the phenomenological models as given in ~\citet{Margueron2018}. }
    
    \label{tab:prior_posterior}
\end{table*}

\begin{figure*}
    \centering
    \includegraphics[width=\linewidth]{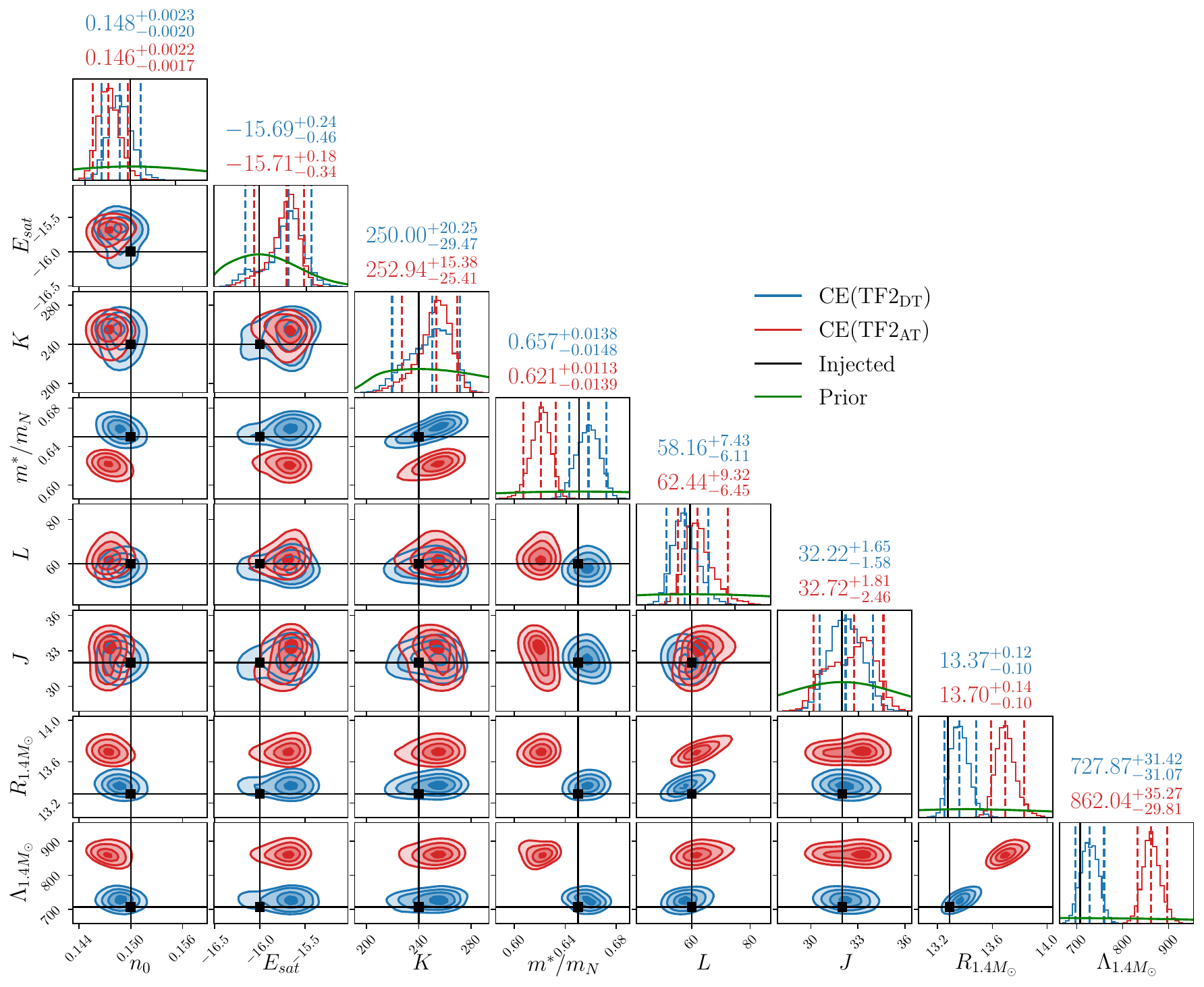}
    \caption{Same as ~\cref{fig:Nuclear_aplus} but with CE detector configuration.}
    \label{fig:Nuclear_CE}
\end{figure*}

We display the joint posteriors of the recovered nuclear parameters in~\cref{fig:Nuclear_aplus} and~\cref{fig:Nuclear_CE} from 13 BNS events with A+ and CE configurations, respectively. We have tabulated the median and 90\% symmetric CI of the EOS parameters recovered with both $\rm TF2_{AT}$ and $\rm TF2_{DT}$ waveform models in ~\cref{tab:prior_posterior}. In addition to the nuclear parameters, we have also shown the distributions of radius ($R_{1.4M_{\odot}}$) and tidal deformability  ($\Lambda_{1.4M_{\odot}}$) of a canonical $1.4M_{\odot}$ NS star reconstructed from the nuclear parameters resulting from different network configurations and different GW waveform models in ~\cref{fig:Nuclear_aplus,fig:Nuclear_CE}. Although we perform the parameter estimation of nuclear parameters, we have shown the equivalent priors of $R_{1.4M_{\odot}}$, and $\Lambda_{1.4M_{\odot}}$ in ~\cref{fig:Nuclear_aplus,fig:Nuclear_CE} corresponding to the priors of the nuclear parameters. From ~\cref{fig:Nuclear_aplus}, one can conclude that with a set of BNS events detected in A+, the nuclear parameters get well-constrained, and especially the determination of nuclear parameters $n_0$, $L$,  and $J$ improves significantly with consideration of BNS events with CE configuration. Additionally, $m^*$ gets significantly constrained to 5\%  with A+, and the uncertainty is further reduced to 3\%  considering CE network configuration. A better constraint on $m^*$ ($\Delta m^*_{\rm 90\%}$ $\leq$ 5\%) is expected, as, in the considered RMF model, $m^*$ controls the stiffness of the EOSs as well as shows strong correlations with NS observables~\citep{Hornick,PRADHAN2023122578,Ghosh2022,Pradhan2022}.

 Furthermore, one can find the overlapping region for most of the EOS parameters recovered with $\rm TF2_{AT}$ and $\rm TF2_{DT}$. However,  the distribution of $m^*$  under two different waveform models seems to be distinct. The ignorance of dynamical tide estimates a lower  $m^*$ by $\sim 6\%$ compared to the value of $m^*$ obtained considering the dynamical tidal correction. The ignorance of dynamical tide estimates a higher $\Lambda$ value (see, ~\cite{Pratten2022}) that requires a stiffer EOS. In contrast, $\rm TF2_{DT}$ provides the lower estimate for $\Lambda$, which can be explained with the softer EOSs. 
 Hence, the ignorance of dynamical tide predicts a lower $m^*$ compared to the value of $m^*$ resulting from the inclusion of dynamical tides, such that the posterior produces sufficiently stiff EOSs to explain the higher $\Lambda$ values. For completeness, we have also displayed the bias on  EOS and  $M-R$ relations recovered from the posteriors of nuclear parameters in Appendix~\ref{app:EOS_MR}. The trend in $M-R$ relation and the properties of $1.4M_{\odot}$ due to ignorance of dynamical tide resulting with the considered RMF model seems to be in excellent agreement with the results from ~\cite{Pratten2022} \footnote{Polytropic EOS model is considered to describe the NS matter}.  


\section{Discussions}\label{sec:conclusion}

We analyze the impact of the inclusion of f-mode dynamical tidal correction in the GW model on the nuclear parameters inferred from GW observations. For the BNS event GW170817, we do not find any substantial impact on the nuclear parameters due to consideration of dynamical tidal correction apart from a smaller lower bound on $m^*$ due to the ignorance of dynamical tide. For future events, all nuclear parameters other than $m^*$ are recovered well even after ignorance of f-mode dynamical tides. With consideration of an ensemble of BNS events in A+, the ignorance of f-mode dynamical corrections lowers the median of $m^*$ ($\sim6\%$) compared to the median of $m^*$ recovered with consideration of f-mode dynamical tide. In contrast, the value of $m^*$ is constrained up to 5\% (within 90\%  CI). Furthermore, with the simulated BNS events in the CE network configuration,  $m^*$ gets constrained up to $\sim 3\%$ ($\leq 90\%$ CI) with a bias of $\sim 6\%$ on the median due to ignorance of dynamical tidal correction. 
We have also discussed how well the other nuclear parameters, such as $n_0$,$E_{sat}$, $K$, $J$, and $L$  get constrained with an ensemble of BNS events in the framework of the RMF model. 
Any biases in these GW measurements can directly impact our understanding of nuclear physics. Therefore, having an accurate description of the GW waveform model is one of the key requirements. Notably, our work sheds light on the impact of dynamical tidal corrections on nuclear physics, especially when constrained by GW observations. This contribution holds significant importance in revealing these effects and advancing our understanding of the intricate interplay between GW astronomy and nuclear physics.

This work can be further improved by considering the additional corrections, such as spin and eccentricity, to the GW model considered in Sec.\ref{sec:injection} for future events. It has been discussed that the consideration of spin and eccentricity further enhances the excitation of oscillation modes of NSs ~\citep{Kuan2022,Steinhoff2021,Pnigouras2022} and may affect GW, and therefore need to be implemented. The EOS model for connecting the nuclear parameter and NS matter is specific to a nucleonic RMF model. Further studies can be performed by considering different NS matter compositions or with other EOS models connecting to nuclear and hypernuclear parameters. 

In this work, we assume the mass distribution of BNS is precisely known, which is not the case in the real data. Furthermore, it is essential to infer NS EOS and mass distribution simultaneously to alleviate the biases arising from individual analyses~\citep{Wysocki:2020myz, Golomb:2021tll}. The Bayesian formalism employed in this work can be further improved by incorporating the cosmological parameters~\cite {Ghosh:2022muc}. However, the most favorable approach would involve the joint fitting of astrophysical distribution as well as cosmological parameters for accurate inference of NS EOS~\citep{dark_bns}. Additionally, it may be essential to incorporate the spin distribution of BNS as well~\citep{Biscoveanu:2021eht}, which needs to be studied in the future. {Further improvements in constraining the NS EOS can be achieved by incorporating joint information from existing electromagnetic (EM) observations and future observations, including those from space X-ray and $\gamma$-ray observatories, along with GW observations~\footnote{The constraints on the NS properties such as the moment of inertia and NS radius in light of future NS measurements with   GW and X-ray observations has been addressed recently in ~\citep{Suleiman2024}.}.}

\section{Acknowledgements}
The authors are thankful to Geraint Pratten for carefully reviewing the manuscript and providing useful suggestions. B.K.P. would like to thank Aditya Vijaykumar for the helpful discussion on the GW waveform model and its implementation for GW parameter estimation. The authors gratefully acknowledge the use of the IUCAA LDG cluster Sarathi, accessed through the LIGO-Virgo-KAGRA Collaboration. B.K.P. is also grateful for the computational resource Pegasus, provided by IUCAA for the computing facility for the computational/numerical work. This material is based upon work supported by NSF's LIGO Laboratory, which is a major facility fully funded by the National Science Foundation. This work makes use of \texttt{NumPy} \citep{NumPy}, \texttt{SciPy} \citep{Scipy}, \texttt{astropy} \citep{2013A&A...558A..33A, 2018AJ....156..123A,astropy2022}, \texttt{Matplotlib} \citep{Hunter:2007}, \texttt{jupyter} \citep{jupyter}, and  \texttt{pandas} \citep{mckinney-proc-scipy-2010} software packages. 
This document has been assigned to the LIGO document number LIGO-P2300407.

\appendix

\section{Distribution of $m^*$ and $L$ for GW170817}\label{app:mstar_LwithGW170817} 
\begin{figure}
     \centering
     \includegraphics[width=\linewidth]{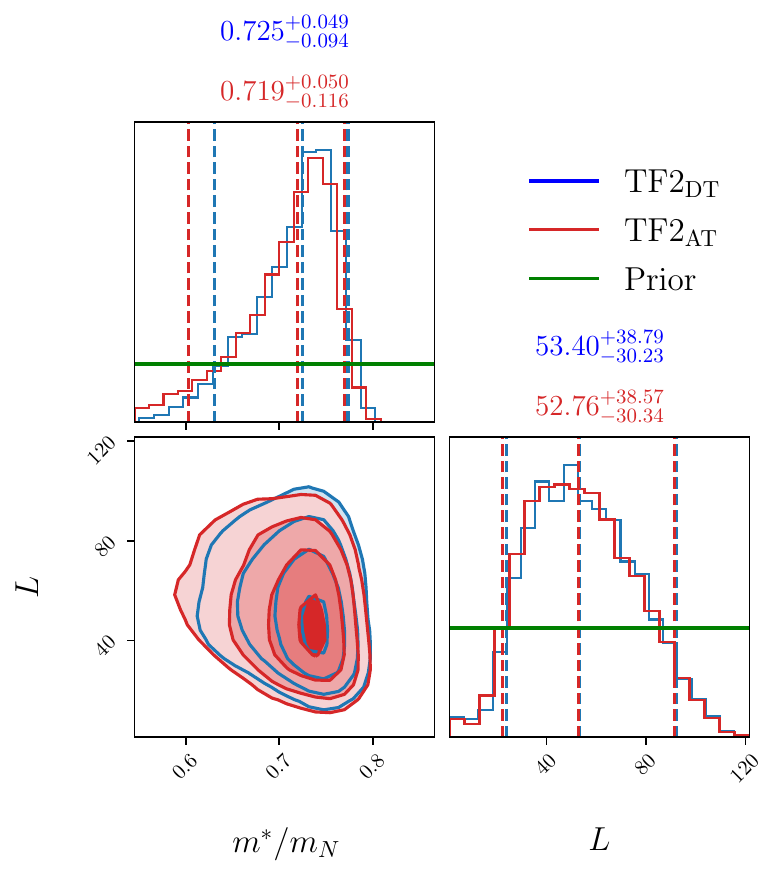}
     \caption{The joint and marginalized posterior distribution of $m^*$ (in units of nucleon mass $m_N$) and $L$ (in MeV) resulting from the BNS event Gw170817 with (blue) and without (red) consideration of f-mode dynamical tidal correction in the GW model.}
     \label{fig:gw170817_mstar_lsym}
 \end{figure}

We display the joint and the marginalized distribution of $m^*$ and $L$ recovered for the BNS event GW170817 analyzed with the GW waveform model $\rm TF2_{AT}$ and $\rm TF2_{DT}$ in ~\cref{fig:gw170817_mstar_lsym}. We do not notice any significant impact of the dynamical tidal correction on the nuclear parameters for the event GW170817. 

\section{Uncertainty in EOS and $M-R$ from Future Events}\label{app:EOS_MR}
\begin{figure*}
\begin{center} 

\subfigure[]{%
\centering
  \includegraphics[width=0.45\textwidth,height=0.33\textwidth]{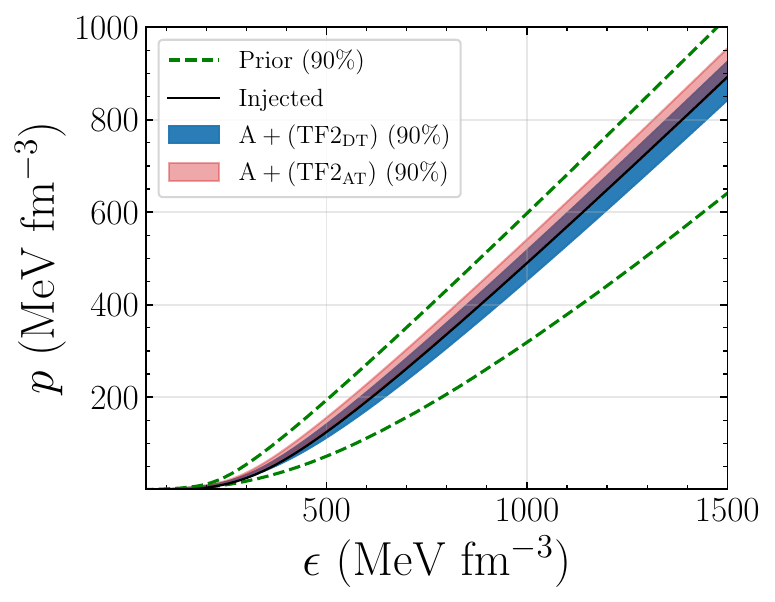}%
  \label{fig:EOS_properties_Aplus}%
}
\subfigure[]{%
\centering
  \includegraphics[width=0.45\textwidth,height=0.33\textwidth]{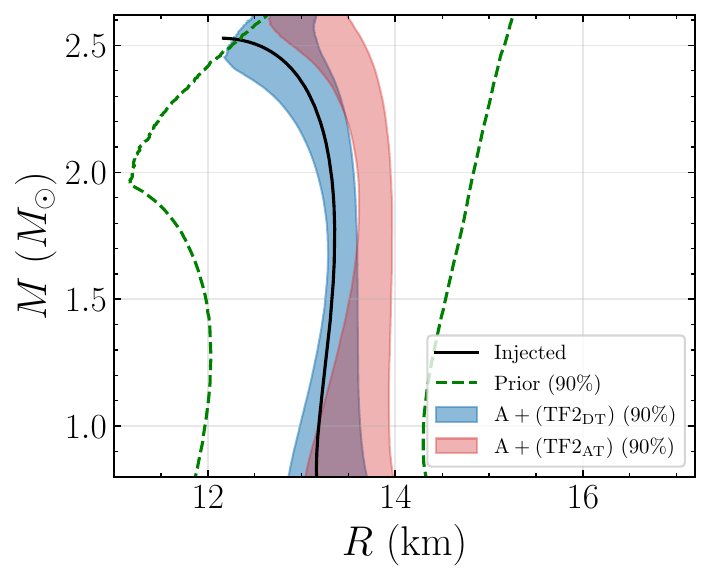}
    \label{fig:MR_properties_aplus}
}
\end{center}
\caption{ (a) 90\% CI for pressure as a function of energy density and (b) 90\% CI for radius as a function of mass reconstructed using the recovered nuclear parameters from the detection of 13 events in A+ network configuration, with (blue) and without (red) consideration of dynamical tidal correction in the recover GW model.  }
\label{fig:EOS_MR_aplus}
\end{figure*}

\begin{figure*}
\begin{center} 

\subfigure[]{%
\centering
  \includegraphics[width=0.45\textwidth,height=0.33\textwidth]{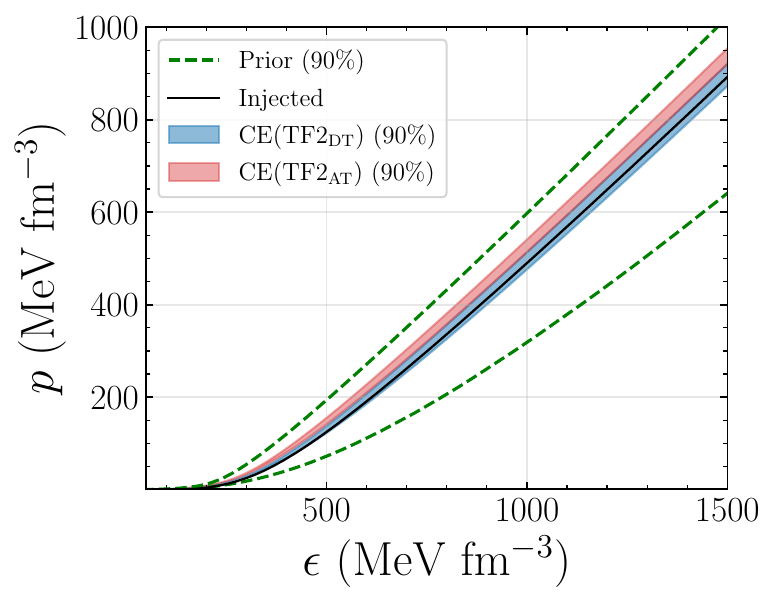}%
  \label{fig:EOS_properties_CE}%
}
\subfigure[]{%
\centering
  \includegraphics[width=0.45\textwidth,height=0.33\textwidth]{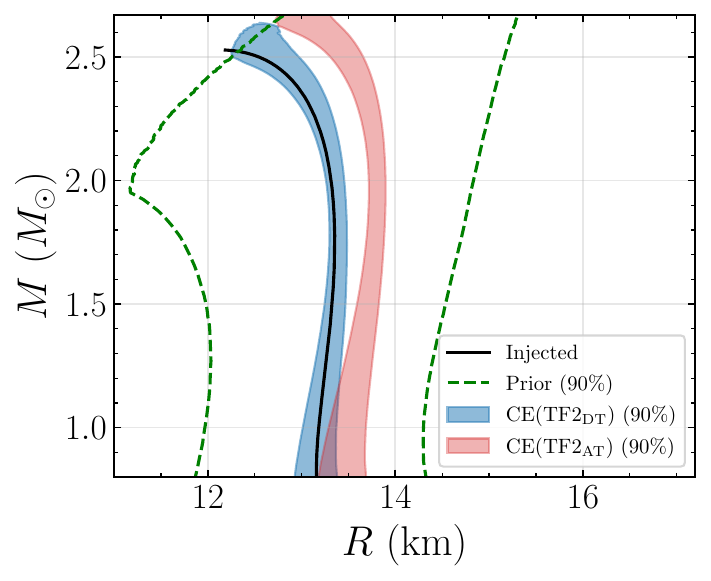}
    \label{fig:MR_properties_CE}
}
\end{center}
\caption{Same as ~\cref{fig:EOS_MR_aplus} but with CE detector network configuration. }
\end{figure*}

 We reconstruct the NS EOS and the $M-R$ posterior from the nuclear parameters recovered using 13 BNS events in A+ and displayed in ~\cref{fig:EOS_properties_Aplus,fig:MR_properties_aplus}, respectively. The substantial bias on the $M-R$ recovered RMF model (shown in  ~\cref{fig:MR_properties_aplus})  due to ignorance of the f-mode dynamical tides is in agreement with the results of ~\cite{Pratten2020}, where a polytropic description of NS matter is adopted. The uncertainty on the NS EOS and $M-R$ relations resulting from the posterior of nuclear parameters recovered with BNS events in CE configuration is shown in ~\cref{fig:EOS_properties_CE,fig:MR_properties_CE}, respectively. As expected, with the increasing sensitivity of the CE detectors,  the uncertainty on EOS or $M-R$ decreases significantly compared to what is obtained with A+. 

\bibliography{Pradhan}{}
\bibliographystyle{aasjournal}



\end{document}